\documentstyle[12pt]{article}
\newcommand{\la}{\label}

\newcommand{\bfx}{{\bf \vec x}}
\newcommand{\bfn}{{\bf \vec n}}

\newcommand{\bfB}{{\bf \vec B}}
\newcommand{\bfv}{{\bf \vec v}}
\newcommand{\bfU}{{\bf \vec U}}
\newcommand{\bfp}{{\bf \vec p}}
\newcommand{\bfA}{{\bf \vec A}}

\newcommand{\be}{\begin{equation}}
\newcommand{\ee}{\end{equation}}
\newcommand{\ba}{\begin{eqnarray}}
\newcommand{\ea}{\end{eqnarray}}
\newcommand{\bastar}{\begin{eqnarray*}}
\newcommand{\eastar}{\end{eqnarray*}}
\newcommand{\half}{{1 \over 2}}

\begin{document}
\begin{titlepage}
 
\vskip 2.0truecm
\begin{center}
$ ~$
\end{center}
\begin{center}
{ 
\bf \large \bf MAGNETIC GEOMETRY AND THE CONFINEMENT 
\\ \vskip 0.2cm
OF ELECTRICALLY CONDUCTING PLASMAS \\
}
\end{center}

\vskip 0.9cm

\begin{center}
{\bf L. Faddeev$^{\, * \, \sharp}$ } {\bf \ and \ } {\bf
Antti J. Niemi$^{\, ** \, \sharp}$ } \\
\vskip 0.6cm

{\it $^*$St.Petersburg Branch of Steklov Mathematical
Institute \\
Russian Academy  of Sciences, Fontanka 27 , St.Petersburg, 
Russia$^{\, \ddagger}$ } \\

\vskip 0.3cm

{\it $^{**}$Department of Theoretical Physics,
Uppsala University \\
P.O. Box 803, S-75108, Uppsala, Sweden$^{\, \ddagger}$} \\

\vskip 0.3cm
and \\
\vskip 0.3cm

{\it $^{\sharp}$Helsinki Institute of Physics \\
P.O. Box 9, FIN-00014 University of Helsinki, Finland} \\
\end{center}

\vskip 1.5cm
We develop an effective field theory approach to inspect
the electromagnetic interactions in an electrically neutral 
plasma, with an equal number of negative and positive charge 
carriers. We argue that the static equilibrium configurations 
within the plasma are topologically stable solitons, 
that describe knotted and linked fluxtubes of helical magnetic 
fields. 
\vfill

\begin{flushleft}
\rule{5.1 in}{.007 in} \\
$^{\ddagger}$  \small permanent address \\ \vskip 0.2cm
$^{*}$ \hskip 0.2cm {\small  E-mail: \scriptsize
\bf FADDEEV@PDMI.RAS.RU and FADDEEV@PHCU.HELSINKI.FI } \\
$^{**}$  {\small E-mail: \scriptsize
\bf NIEMI@TEORFYS.UU.SE}  \\
\end{flushleft}
 
\end{titlepage}

\vskip 0.7cm
\rm
\noindent
Plasma comprises over 99.9 per cent of known matter in 
the Universe. However, among the different states of matter 
its physical properties are the least understood. This is
largely due to a highly complex and nonlinear behaviour,
which makes theoretical investigations quite difficult.
Particularly notorious are the instabilities 
that hamper plasma confinement in thermonuclear fusion 
energy experiments \cite{freid}.

In the present Letter we consider the electromagnetic
interactions within a charge neutral plasma,
with an equal number of negative and positive
charge carriers. We propose a first principles
field theory model to describe the fluid dynamical
properties of this plasma, and find results that 
challenge certain widely held views on plasma behaviour. 
In particular, we argue that stable self-confining 
plasma filaments can exist, and are described by 
topologically nontrivial knotted solitons. 

In magnetohydrodynamics \cite{freid} the geometrical
properties of an electrically neutral plasma 
are conventionally described using a single-fluid 
approximation. The individual charged particles 
contribution is described collectively by the hydrostatic
pressure $p$, which according to standard kinetic 
theory relates to the kinetic energies of the 
individual particles $p \propto m v^2$. The equation
of motion then follows from the properties of the
pertinent energy-momentum tensor $T_{\mu\nu}$,
the spatial part of its divergence coincides with
the external dissipative force which leads to 
the Navier-Stokes equation
\be
\rho \frac{d \bfU}{dt} \ = \ - \nabla p \ + \ (\nabla \times
\bfB) \times \bfB \ + \ \eta \nabla^2 \bfU^2
\la{navier}
\ee
Here $\bfU$ is the bulk (center of mass) velocity 
of the plasma, and $\eta$ is the coefficient of viscosity.  
The plasma evolves according to (\ref{navier}), 
dissipating its kinetic energy by the viscous force. 
This force is present whenever the plasma is in 
motion but ceases when the plasma reaches a magnetostatic 
equilibrium configuration. In that limit the Navier-Stokes
equation reduces to a balance relation between the 
gradient of the hydrostatic pressure and the magnetic force,
\[
\nabla p \ = \ (\nabla \times \bfB) \times \bfB
\] 
Ideally, one might expect that under proper conditions a plasma 
in isolation becomes self-confined due to the currents that flow
entirely within the plasma itself. But this appears to be excluded
by a simple virial theorem \cite{freid} which suggests that any
static plasma configuration in isolation is dissipative.
As a consequence of such apparently inborn instabilities,
strong external currents are then commonly introduced to
confine a plasma in laboratory experiments.

We now argue that there are important non-linear
effects which are not accounted for by a structureless mean
field variable such as the pressure $p$. These nonlinearities
have their origin in the electromagnetic interactions between
the charged particles within the plasma. They remain
hidden when the energy-momentum tensor relates to the kinetic
energies of the individual particles, but become visible once
we recall the familiar but nontrivial relation between 
the kinetic momentum $m\bfv$ and the canonical 
momentum $\bfp$ of a charged point particle, 
$$ 
m \bfv \ = \ \bfp - e \bfA
$$ 
where $\bfA$ is the electromagnetic vector potential. We
propose that when these electromagnetic forces within
the plasma are properly accounted for, the ensuing field
theory model has the potential of supporting stable
soliton-like configurations which describe helical, 
self-confined structures within the plasma medium.

Our starting point is a natural kinetic field theory 
model of a two-component plasma of electromagnetically
interacting charged point particles such as electrons and
deuterons.  In natural units the classical action is
\[
S \ = \ \int dt d^3x \ \biggl[ \ i {\psi_e}^* (\partial_t +
i e A_t) \psi_e \ + \ i {\psi_i}^* (\partial_t -
i e A_t) \psi_i \ - \  \frac{1}{2m} | (\partial_k +
i e A_k) \psi_e |^2
\]
\be
- \ \frac{1}{2M} | (\partial_k -
i e A_k) \psi_i |^2 \ - \ \frac{1}{4} {F_{\mu\nu}^2}
\biggr]
\la{act}
\ee
As usual $F_{\mu\nu} = \partial_\mu A_\nu -
\partial_\nu A_\mu$. The $\psi_e$ and $\psi_i$ are two (complex)
non-relativistic fields for electrons and ions with masses $m$ 
and $M$ and electric charges $\pm e$, respectively. Notice that we 
describe both charged fields by macroscopic (Hartree-Fock)
wave functions. This is adequate in the classical
Bolzmannian limit which is relevant 
in conventional plasma scenarios \cite{freid}. 
The action (\ref{act}) determines our first principles 
description of a non-relativistic
plasma. Its magnetohydrodynamical properties are
governed by the pertinent energy-momentum tensor $T_{\mu\nu}$,
which can be constructed from (\ref{act}) in a standard 
manner. When we include the contributions that account for 
the bulk motion of the plasma medium, this leads to an
appropriate version of the Navier-Stokes equation (\ref{navier}). 
Here we are interested in the ensuing static equilibrium
configurations. These configurations are local minima of the
internal energy $E$, which is determined by the temporal $T_{00}$
component of the energy-momentum tensor. For a stationary plasma
fluid (\ref{act}) we get from (\ref{act})
\[
E \ = \ \int d^3 x \ \biggl[ \
\frac{1}{2\mu} \biggl\{ ~ \sin^2 \!\! \alpha \, |
(\partial_k + i e A_k)
\psi_e |^2 \ + \ \cos^2 \!\! \alpha \, | (\partial_k -
i e A_k) \psi_i |^2 ~ \biggr\}
\]
\be
+ \ \frac{1}{2} B_{i}^2 \ + \ g \bigl( {\psi_e}^* \psi_e -
{\psi_i}^* \psi_i\bigr)^2 \ \biggr]
\la{ene}
\ee
Here $\mu = m \cdot \sin^2 \!\! \alpha 
= M \cdot \cos^2 \!\! \alpha$ is the reduced mass
and $B_i = \half \epsilon_{ijk}F_{jk}$ is the magnetic
field. The quartic potential is the remnant of 
the Coulomb interaction with $g$ an effective
coupling constant. It emerges when 
we first use Gauss' law to eliminate the electric field, and 
then recall that in any realistic plasma the Debye screening 
radius is small in comparison to any characteristic 
length scale of interest.

The free energy (\ref{ene}) is subjected to the conditions that the 
plasma is electrically neutral with an equal (large) 
number $n_e$ of electrons and $n_i$ of ions, $n_e = n_i$
and the total number of charge carriers in the volume $V$
remains intact $n_e + n_i = N$. These conditions can be
implemented by adding appropriate chemical potential terms
to (\ref{ene}) in the usual fashion. But for simplicity
we here account for them as constraints, 
imposed by appropriate boundary conditions.  Besides the terms
that we have displayed in (\ref{ene}) there can also be additional
interaction terms for the charged fields. Such terms are usually 
induced by thermal fluctuations and finite density effects, or by gravitational interactions. However, according to standard 
universality arguments we expect the main features
of (\ref{ene}) to persist at temperatures and distance
scales which are relevant in conventional plasma scenarios.

We propose that (\ref{ene}) yields an adequate
approximation for a non-relativistic plasma in a
kinetic regime where the thermal energy is
sufficiently high to prevent the formation of 
charge neutral bound states, which correspond
to hydrogen atoms in the case of electrons and deuterons. 
Such bound states are present at lower temperatures,
and their presence can be accounted for by terms of the form
\[
E_{bs} \ = \ \int d^3 x \ \biggl[ \ \frac{1}{2}\cdot
\frac{1}{m+M} ( 
\partial_k \Phi )^2 \ + \ \lambda \cdot
\Phi \psi_e \psi_i \ + \ \bar \lambda \cdot 
\Phi \psi^*_e \psi^*_i
\ \biggr]
\]
Here $\Phi$ a real scalar field that describes 
a charge neutral bound state of $\psi_e$ and $\psi_i$.
At a sufficiently high temperature this bound
state degree of freedom decouples, and (\ref{ene})
becomes adequate for describing the bulk 
properties of the plasma.

Since $n_e = n_i$ we have overall charge neutrality. 
However, there can be local charge density fluctuations 
that should not be ignored. Indeed, we now 
proceed to argue that static charge density 
fluctuations are naturally present in (\ref{ene}).
These fluctuations accompany stable, static solitons 
which describe filamental self-confined structures within 
the plasma. For this we first note that the different 
contributions in (\ref{ene})
respond differently to a scaling $\bfx \to \lambda \bfx$.
The kinetic terms scale in proportion to $\lambda$ and
the Coulomb potential in proportion
to $\lambda^3$, but the magnetic energy scales
like $\lambda^{-1}$. Consequently the
existence of nontrivial, non-dissipative plasma
configurations in (\ref{ene}) can not be excluded
by simple virial arguments, quantitative investigations
become necessary.

We start by observing that the vector
potential $A_k$ enters at most quadratically. Consequently
it can be eliminated: We vary (\ref{ene}) {\it
w.r.t.} $A_k$ and get
\[
A_k \ = \  \frac{1}{2 e}\cdot \frac{1}{\sin^2\! \alpha \,
|\psi_e|^2 + \cos^2 \! \alpha \,  |\psi_i|^2}
\biggl[ \, i\sin^2 \! \alpha \cdot
({\psi_e}^* \partial_k \psi_e - \partial_k {\psi_e}^* \psi_e) \]
\be
-i \cos^2 \! \alpha \cdot ( {\psi_i}^* \partial_k \psi_i -
\partial_k {\psi_i}^* \psi_i ) \ - \
\frac{2\mu}{e} \cdot \epsilon_{kij}\partial_i
B_{j} \biggr]
\la{A}
\ee
which determines $A_k$ in terms of an iterative
gradient expansion, in powers of derivatives in the
charged fields.
We introduce new variables by
\be
\bigl( \ \psi_e \ ,  \psi_i \ \bigr) \ = \
\rho \cdot \bigl( \ \cos \alpha \cdot \sin \frac{\theta}{2}
\ e^{i \varphi} \ , \
\sin \alpha \cdot \cos \frac{\theta}{2} \ e^{i\chi} \ \bigr)
\la{Psi}
\ee
For reasons that will soon become obvious we have chosen
these variables so that they are natural for describing
tubular field configurations, with $\varphi$ and $\chi$
related to the toroidal and poloidal angles and $\theta$
a shape function that measures the distance away from the
centerline of the tube. We compute the free energy (\ref{ene})
to the leading order in a self-consistent gradient
expansion, where we keep only terms which are at most
fourth order in the derivatives of
the variables (\ref{Psi}). This approximation is adequate
in conventional plasma scenarios where the fields are
relatively slowly varying. We start by
determining $A_k$ from (\ref{A}) iteratively in the
variables (\ref{Psi}). We substitute the result
in (\ref{ene}), and by defining a three-component
unit vector $\bfn  = ( \cos(\chi+\varphi) \sin
\theta  \, ,  \, \sin(\chi + \varphi)\sin \theta \, , \,
\cos \theta )$ we finally get for the free energy
\be
E \, = \, \int d^3x \biggl[ \, \frac{1}{2} \cdot \frac{1}{m+M}
\cdot \biggl\{ \, (\partial_k \rho)^2 \, +
\, \rho^2 \cdot |\partial_k \bfn |^2 \, \biggr\}
\, + \, \frac{1}{4e^2} (\bfn \cdot \partial_i
\bfn \times \partial_j\bfn
)^2 \, + \, \frac{g\rho^4}{4} ( n_3 -
\cos 2\alpha )^2 \, \biggr]
\la{fadd}
\ee
We note that since $m$ and $M$ are
both nonvanishing, overall charge
neutrality implies that asymptotically $\theta \to
2 \alpha \not= n\pi$. Since $\rho \rightarrow \, 
const. \, \not= 0$
asymptotically (see below),
the Coulomb interaction then yields a mass term
for the variable $\theta$. We also note that (\ref{fadd})
naturally embodies a helical structure, described by the
Hopf invariant \cite{nature}. To the relevant order in
our gradient expansion
\be
Q_H \ = \  -\frac{1}{e^2 4\pi^2} \int d^3x \ \bfB \cdot \bfA \ = \
-\int d^3x \ \nabla \! \cos \theta \cdot
\frac{\nabla \varphi}{2\pi} \times
\frac{\nabla \chi}{2\pi} \ = \ \Delta \varphi\cdot \Delta
\chi
\la{hopf}
\ee
Here $\Delta \varphi$ {\it resp.} $\Delta \chi$ denotes the
($2\pi$) change in the pertinent variable over the (would-be)
tube, when we cover it once in the toroidal and poloidal
directions over a magnetic flux surface with constant $\theta$.

The field $\rho$ is a measure of the particle
density in the bulk of the plasma. If its average (asymptotic)
value $<\!\!\!\rho^2\!\!\! > = \rho_0^2$ becomes too small, 
the collective behaviour of the plasma will be lost and
instead we have an individual-particle behaviour 
of the charged constituents, interacting via Coulomb collisions. 
Consequently we select the average $\rho_0^2$ so that it 
acquires a sufficiently large value in the medium.
Local charge fluctuations then occur in regions
where the unit vector $\bfn$ becomes a variable so that
$\theta \not= 2\alpha$. According to our
adiabatic approximation $|\partial_k  \bfn |$ is a
slowly varying bounded function over the entire
charge fluctuation region, and in particular it
vanishes outside of the fluctuation region.
When we inspect the $\rho$-equation of
motion that follows from (\ref{fadd}) we find that
it can be related to a Schroedinger equation for the
lowest energy scattering state in an
external potential $ \propto |\partial_k  \bfn |^2$. From 
this we then conclude that $|\rho(\bfx)|$
never vanishes; it is bounded from below by a
non-vanishing positive value which is related to the 
ensuing scattering length.
This implies that if we average the free energy (\ref{fadd})
over $\rho(\bfx)$, to the relevant order in our gradient
expansion the result can be related to the universality
class determined by the Hamiltonian
\be
H \ = \ \int d^3x \ \biggl[
\ \gamma \cdot |\partial_k \bfn |^2  \ + \ \frac{1}{4e^2}
(\bfn \cdot \partial_i \bfn \times \partial_j\bfn
)^2 \ + \ \lambda \cdot ( n_3 - \cos 2\alpha )^2 \, \biggr]
\la{fad2}
\ee
where $\gamma, \lambda$ are nonvanishing
positive constants, proportional to the scattering length
of our Schroedinger equation. This Hamiltonian is known to support
stable knotlike solitons \cite{nature}. In particular,
since the third (Coulomb) term is positive it does
not interfere with the lower bound estimate
derived in \cite{vak}. This estimate states that the first two
terms in (\ref{fad2}) are bounded from below by the 
fractional power $|Q_H|^{3/4}$ of the Hopf invariant. 
Even though we do not expect that in the case
of (\ref{fadd}) this lower bound estimate 
remains valid as such,
we nevertheless conclude that when $Q_H \not=0$ the energy
(\ref{fadd}) admits a nontrivial lower bound;
the conclusions from the virial theorem
in \cite{freid} should not be adapted too hastily.

The properties of (\ref{fad2}) with $\lambda=0$
have been studied in \cite{nature}-\cite{hie}.
In particular, the  numerical simulations
in \cite{sut}, \cite{hie} clearly confirm
the existence of stable, knotted and linked solitons
with a nontrivial Hopf invariant \cite{nature}.
The present considerations
firmly suggest that the conclusions in \cite{nature}-\cite{hie}
prevail also in the case of (\ref{fadd}). Indeed,
we have tentatively verified that similar solitons are
present in (\ref{fadd}), by numerically
constructing a line vortex soliton in this model; we
describe our solution in figure 1. These solitons then become natural
candidates for describing filamental and toroidal
structures in the plasma, including
coronal loops above the solar photosphere and the
design of magnetic geometries in thermonuclear fusion
energy experiments. The numerical simulations reported
in \cite{sut}-\cite{mauri} are very extensive, and clearly
reveal the complexity of the problem. Accordingly
the interest has thus far mainly
concentrated on the identification of soliton geometries,
very little is still known about the solitons detailed
physical properties. Consequently at this time
we are not in a position to present definite physical
predictions in the context of actual applications, high
precision numerical methods still remain under active
development \cite{sut}, \cite{hie} and
we have to limit ourselves to a few general remarks:
In the numerical simulations that have been completed
thus far, it has been found that for generic
integer values $(\Delta \varphi,
\Delta \chi) = (n,m)$ in (\ref{hopf}) the $\lambda = 0$
solitons of (\ref{fad2}) form involved knotted and
linked structures. Such complex geometries might be
natural in a number of applications, for example when
modelling coronal loops. But they might not be of any
immediate practical interest for the design of plasma
geometries in fusion energy experiments, where planar
toroidal configurations are preferable. Indeed, there 
are also a few torus-shaped solitons which are
essentially planar. These occur for values $(n,m)
= (1,1), (2,1), (1,2), (2,2)$ \cite{hie}. The
simplest one is $(1,1)$ but it appears
to have an energy density that peaks at the toroidal
symmetry axis. As such this may be an advantage
in designing actual fusion reactors. But it 
could also become problematic,
as it may interfere with the construction of an
external torus-shaped coiling system which should be needed to
create the soliton. On the other hand, the $(2,1)$
soliton seems to have a torus-shaped
energy density distribution which vanishes
at the symmetry axis and peaks at the
centerline of the torus (see \cite{hie}). 
Since this soliton is also quite sturdy \cite{hie}, 
it is a natural candidate {\it e.g.}
for designing magnetic geometries for thermonuclear
fusion energy purposes. In particular, this configuration 
strongly suggests that for a stable, toroidal planar geometry 
the safety factor \cite{freid} in the bulk
of the plasma should not exceed $q\approx 2$.
A configuration with a higher value for $q$ tends to adjust itself towards a geometrical shape which is
not planar; see the computer animations in the www-address of 
reference \cite{hie}. 

\vskip 0.5cm
\noindent
In conclusion, we have argued that an electrically neutral
conducting plasma can form stable, self-confining structures.
This is due to soliton-like solutions, which we have shown
will appear when we properly account for the nontrivial
electromagnetic interactions within the plasma. We
have proposed that our solitons can become relevant
in a number of practical scenarios, including coronal loops and
the design of magnetic geometries in thermonuclear 
fusion energy experiments. However,
in order to assess the impact of our findings,
detailed numerical investigations are necessary.
Unfortunately the simulations remain highly complex,
even with the present day supercomputers. Consequently
we have not been able to reliably confirm that parameters
such as the asymptotic density $\rho_0$ and the coupling $g$
can indeed be selected appropriately for the solitons to have
direct technological relevance for example in the design 
of magnetic geometries for energy producing thermonuclear fusion 
reactors. But  
since over 99.9 per cent of all known 
matter in the Universe exists in the plasma state, there are 
no doubt numerous scenarios where our results can 
become important. Besides astrophysical applications or quark-gluon
plasma experiments, these might include even an explanation to
the highly elusive ball lightning.

\vskip 1.0cm

We thank A. Alekseev, E. Babaev, A. Bondeson,
H. Hansson, E. Langmann, V. Maslov, H.K. Moffatt, S. Nasir, 
A. Polychronakos, R. Ricca and
G. Semenoff for discussions. We are particularly indebted to
M. L\"ubcke for his help, and to J. Hietarinta
for communicating the results in \cite{hie}
prior to publication. We thank the Center for
Scientific Computing in Espoo, Finland for the use of
their computers. The work of L.F. has been 
supported by grants RFFR 99-01-00101 and INTAS 9606, and
the work of A.J.N. has been supported by NFR Grant F-AA/FU 
06821-308.

\vfill\eject

\vfill\eject
\begin{flushleft}
{\bf Figure Caption}
\end{flushleft}
\vskip 1.0cm


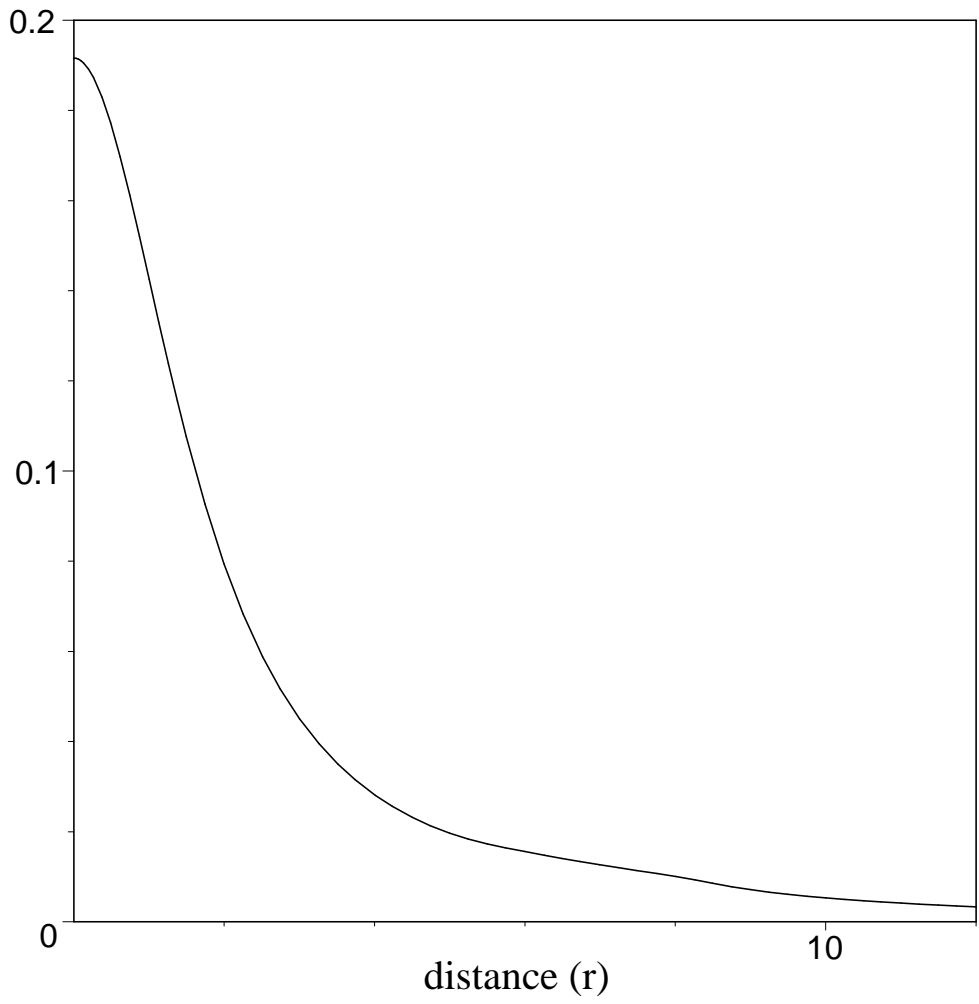
{\bf figure 1:} An example of a numerically constructed
tubular line vortex solution of (\ref{fadd}), with energy
density plotted as a function of the distance from the tubular
center-line. We use standard cylindrical
coordinates $(r,\phi,z)$ so that the tubular center-line
coincides with the $z$-axis. For simplicity we have taken
a limit of large ion mass which sends $2 \alpha \to \pi$.
All numerical parameters in (\ref{fadd}) are ${\cal O}(1)$
and the helical structure is characterized by $\varphi
+ \chi \, = \, \phi + 0.6 \, z$.

\end{document}